\documentclass[aps,prl,twocolumn,amsmath,amssymb,10pt]{revtex4}
\usepackage{graphicx}
\usepackage{ulem}
\usepackage{soul}
\usepackage{txfonts}
\usepackage[colorlinks, linkcolor=blue]{hyperref}
\usepackage{verbatim}
\usepackage{esint}
\renewcommand{\vec}[1]{\boldsymbol{#1}}

\def \n {{\vec n}}

\def \n{{\mathbf{n}}}

\def \beq {\begin{eqnarray}}
\def \eeq {\end{eqnarray}}
\def \tn {\textnormal}

\def \vp {\varphi}
\begin{document}

\title{Slow scrambling and hidden integrability in a random rotor model} 

\author{Dan Mao}
\author{Debanjan Chowdhury}
\author{T. Senthil}
\affiliation{Department of Physics, Massachusetts Institute of Technology, Cambridge Massachusetts
02139, USA.}
\begin{abstract}
We analyze the out-of-time-order correlation functions of a solvable model of a large number, $N$, of $M$-component quantum rotors coupled by Gaussian-distributed random, infinite-range exchange interactions. We focus on the growth of commutators of operators at a temperature $T$ above the zero temperature quantum critical point separating the spin-glass and paramagnetic phases. In the large $N,~M$ limit, the squared commutators of the rotor fields do not display any exponential growth of commutators, in spite of the absence of any sharp quasiparticle-like excitations in the disorder-averaged theory. We show that in this limit, the problem is integrable and point out interesting connections to random-matrix theory. At leading order in $1/M$, there are no modifications to the critical behavior but an irrelevant term in the fixed-point action leads to a small exponential growth of the squared commutator. We also introduce and comment on a generalized model involving $p$-pair rotor interactions.   
\end{abstract}
\maketitle

\setcounter{tocdepth}{2}

{\it Introduction.-} The route to thermalization \cite{Deutsch,Srednicki,Tasaki,rigol}, or lack thereof \cite{PWA58,FA80}, in closed quantum systems remains one of the central open questions in quantum statistical mechanics. The problem has been studied extensively over the past few decades and has received renewed attention in recent years \cite{Huserev,Altmanrev}, partly as a result of great advances in experimental techniques that allow us to probe quantum dynamics in closed systems \cite{Greiner,schmied,Blatt14,bloch1,bloch2}. 
The eventual global thermalization of closed quantum systems may, in general, proceed through multiple distinct stages. At the shortest timescales, there is a decay of local perturbations through the process of relaxation and this timescale affects the time dependence of simple two-point correlation functions of local observables \cite{forster}. At intermediate timescales, a characteristically distinct phenomenon of scrambling \cite{sekino} leads to the spreading of information across all of the degrees of freedom of the system. The connection between scrambling and thermalization remains a topic of active research.

It has been suggested that certain non-trivial out of time ordered correlation (OTOC) functions \cite{LO69} of local operators (defined in Eq.~\ref{ct} below) can probe the onset of scrambling \cite{ShenkerStanford2014,kitaevtalk}. In recent years, these correlation functions have been used extensively to study scrambling in black holes \cite{ShenkerStanford2014,kitaevtalk,Shenker2014}, which are supposed to be the fastest known scramblers \cite{MSS15} in nature. Moreover, since the memory of the initial state is effectively lost \cite{hosur} due to scrambling, it can be heuristically related to an onset of `many-body quantum chaos' \cite{kitaevtalk} with a positive `Lyapunov' exponent. A number of calculations in a variety of models at large $N$ or at weak-coupling have found an exponential growth of the OTOC (see e.g. \cite{Maldacena,DSweak,BSDC17,PatelSachdev,DCBS17,Aleiner16,GuQiStanford,BanerjeeAltman}) but a clear signature of such growth in numerical studies has been absent \cite{Knap16,Luitz,FP18a}. Certain random-circuit models also explicitly point out the absence of such exponential growth of OTOC \cite{AN18,FP18b} but it may be possible to reconcile these differences \cite{Swingle18}. The precise relationship between scrambling and chaos remains unclear.

Inspired by the rapid developments in the study of OTOC in many-body systems, including in the family of Sachdev-Ye-Kitaev (SYK) models \cite{SY92,kitaevkitpsyk,kitaevsuh} which do not admit a description in terms of weakly interacting quasiparticles and scramble at the maximally allowed rate \cite{Maldacena}, we address a number of questions in the remainder of this paper: (i) Are there $(0+1)-$dimensional models different from the SYK model that also scramble at the maximal rate? (ii) Does the absence of well-defined quasiparticle excitations ensure (near-)maximal chaos? (iii) In models with quenched disorder, is it possible to distinguish between the phenomenon of `dephasing' arising as a result of inelastic processes due to strong interactions as opposed to coupling to disorder?  

In this letter we study scrambling in a class of rotor models  \cite{SS1,SS2} that share some superficial resemblance to the SYK models but have remarkably distinct properties. We will begin with a large number, $N$, of $M$ component real rotor fields with $O(M)$ symmetric all-to-all interactions that are assumed to be quenched random variables with zero mean \cite{SS1}. The system can be tuned through a quantum critical point (QCP) separating a paramagnet and a quantum spin glass. Upon disorder averaging, it is known that there are well-defined quasiparticle excitations on either side of the transition but the QCP itself lacks any quasiparticle-like excitations \cite{SS1}. At low energies, the temperature ($T$) serves as the only relevant scale at the QCP \cite{QPT}, and one may then naturally expect the QC regime to scramble with a rate ($\lambda_L$) that is proportional to $(k_BT/\hbar)$ \cite{PatelSachdev,DCBS17}. 

Instead we find results that are surprisingly at odds with this natural expectation. In particular: (i) in the $N,~M\rightarrow\infty$ limit, the OTOC does {\it not} exhibit any exponential growth (i.e. $\lambda_L=0$) in the QC regime, (ii) the model can be solved (in the $N\rightarrow\infty$ limit) to all orders in $1/M$ \cite{SS1,SS2}; the leading (irrelevant) correction in $1/M$ leads to a small $\lambda_L>0$, (iii) the absence of an exponential growth of the OTOC in the former limit, in spite of the apparent non-quasiparticle character, is related to an underlying `hidden' integrability. We demonstrate this by pointing out an interesting connection to a `random-matrix' description \cite{mehta} for the saddle point of the $O(M)$ rotor model at large $N,~M$. Finally, inspired by these new insights, we construct a generalized rotor model with random exchange interactions between $p~(>2)$-pairs of rotors and point out the associated similarities with the generalized SYK$_q$ models with $q~ (>4)$-fermion interactions \cite{sykq}.

{\it Random $O(M)$ rotor model.-} We consider a model of $O(M)$ rotors defined on $N$ sites with random interactions between any two sites. There is no notion of ``space" in this model, which is effectively $(0+1)-$dimensional. The Hamiltonian of the model is given by,
\begin{equation}
H=g\sum_{i}\frac{L_i^2}{2M}+\frac{M}{\sqrt[]{N}}\sum_{i<j} J_{ij}\n_i\cdot \n_j,
\label{ham}
\end{equation}
where $i,j\in \{1,...,N\} $ are site indices, $\n_i$ are $M$ component vectors of unit length, and $L_i$ represent angular momentum tensors with $M(M-1)/2$ components. The components of $\n_i$ are mutually commuting and satisfy standard commutation relations \cite{si}.
The $J_{ij}$ are uncorrelated random variables selected from a Gaussian distribution, $P(J_{ij})=e^{-J_{ij}^2/2J^2}/\sqrt{2\pi J^2}$.

We can obtain the saddle point action in the $N\rightarrow \infty, ~M\rightarrow\infty$ limit, by integrating over disorder configurations \cite{SS1}. In particular, the self-consistency condition for $Q(i\omega_n)$, the Fourier transform of the correlator $Q(\tau)=\langle \n(\tau)\cdot \n(0)\rangle$ (assuming a replica-symmetric solution) is given by \cite{si}
\beq
Q(i\omega_n) = \frac{g}{\omega_n^2 + \lambda - gJ^2 Q(i\omega_n)},
\label{SC}
\eeq
where $\omega_n$ is a Bosonic Matsubara frequency and $\lambda$ is a Lagrangian multiplier that imposes the constraint of unit length for the rotor fields. 

It is immediately clear that the spectral function,  $A(\omega) = \tn{Im}~ Q(i\omega_n\rightarrow\omega+i0^+)$, is given by,
\beq
A(\omega) = \tn{sgn}(\omega) \frac{\sqrt{(\omega^2 - \Delta_s^2)(\Delta_s^2 + 4Jg -\omega^2)}}{2gJ^2}
\eeq
where we have defined $\Delta_s^2 = (\lambda - 2gJ)$, such that $A(\omega)$ is finite if $\Delta_s^2<\omega^2< (\Delta_s^2 + 4Jg)$ and is zero otherwise. 

It is evident then that $\Delta_s$ is the spin-gap to excitations in a paramagnetic phase, which vanishes at the $T=0$ QCP (when $\lambda=2Jg$) between the paramagnet and a spin-glass phase. At the QCP, for frequencies $\omega^2\ll gJ$, $Q_R(\omega)\equiv Q(i\omega_n\rightarrow\omega+i0^+)$ is given by
\beq
Q_R(\omega)\approx -\frac{\omega^2}{2gJ^2} + \frac{1}{J} + i\frac{\tn{sgn}(\omega)}{J\sqrt{gJ}} |\omega|.
\label{qrw}
\eeq
The non-analytic piece in $A(\omega)\sim|\omega|$ (or equivalently, $Q(\tau)\sim1/\tau^2$ for large $\tau$) encodes the non-trivial character of the quantum critical regime. Similar behavior was also obtained for the transverse-field Ising model with random exchange interactions \cite{Husemiller}, which is expected to be in the same universality class as the above model with $M=1$.

The solution of the constraint equation, equivalently expressed as $Q(\tau=0)=1$, can be used to obtain the self-consistently generated `thermal-mass', $\Delta_s(T)$ in the QC region at $g=g_c=9\pi^2J/16$. This leads to \cite{SS1,si}, 
\beq
\Delta_s(T)=\frac{2\pi k_BT}{\sqrt{3}\log^{1/2}(1/T)}.
\label{dst}
\eeq

{\it OTO Correlation functions.-} For any two generic Hermitian operators, $V$ and $W$, we can define an ``unregulated" squared commutator,
\beq
C(t_1,t_2) = \tn{Tr} \{ \rho [V(t_1),W(0)]^\dagger [V(t_2),W(0)]\} ,
\label{ct}
\eeq
where $V(t)=e^{iHt} V e^{-iHt}$ is the time-evolved Heisenberg operator corresponding to the Hamiltonian $H$. The system is taken to be in thermal equilibrium at a temperature $T=\beta^{-1}$ with the density matrix $\rho\propto e^{-\beta H}$.
For the purpose of our computations in this paper, we will instead choose to work with a ``regulated" squared commutator \cite{MSS15} defined in terms of the rotor fields (i.e. $V,~W \equiv \n$),
\beq
F(t_1,t_2)=-\sum_{i,j,a,b}\overline{\tn{Tr}\{\sqrt\rho[n_i^a(t_1) ,n_j^b(0)]\sqrt\rho[n_i^a(t_2) ,n_j^b(0)]}\},
\eeq
where $i,j = 1,..,N$ are site indices, $a,b = 1,..,M$ are the vector indices and ``$^{\overline{~~~}}$" represents disorder averaging. For systems with a large number of degrees of freedom, $N,~M$, the expectation \footnote{Recent work \cite{Galitski} has pointed out that the regulated vs. unregulated OTOC can display $\lambda_L$ that are significantly different from one another; the bound \cite{MSS15} was proposed for the regulated OTOC.} in a non-integrable chaotic system is that
$F(t_1,t_2) \sim \epsilon ~e^{\lambda_L(t_1+t_2)/2}$. Here $\epsilon$ is some small parameter, which may depend on time and on the number of relevant degrees of freedom. 

\begin{figure}[h]
\begin{center}
\includegraphics[scale=0.28]{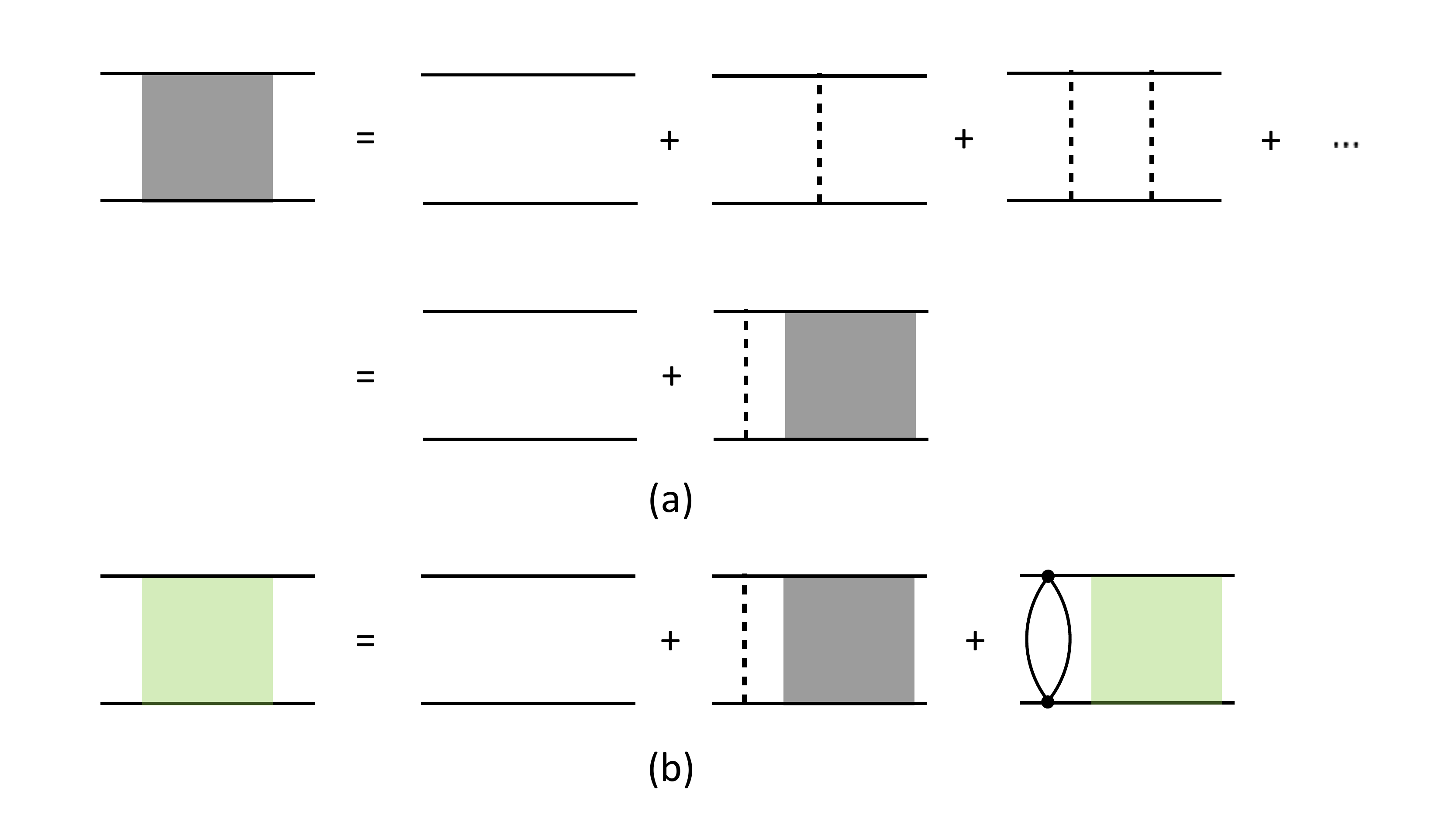}
\end{center}
\caption{Diagrammatic expansion for $F(t_1,t_2)$ (a) in the $N,~M\rightarrow\infty$ limit, and, (b) upon including the leading correction to $O(1/M)$. Solid horizontal (vertical) lines denote the propagator, $Q_R(\omega)$ ($Q_W(\omega)$). Dashed lines denote disorder contraction of $J_{ij}$.}
\label{ladsum}
\end{figure}

For the system described by the Hamiltonian in Eq. \ref{ham}, the leading contribution to $F(t_1,t_2)$ arises from the series of diagrams shown in Fig.\ref{ladsum}(a) in the $N,~M\rightarrow\infty$ limit.
 The two horizontal lines represent the retarded propagators of the $\n$ fields along two real time branches respectively and the vertical dotted lines represent the disorder contraction over the random couplings,  $J_{ij}$. Note that the ladder series is not simply a perturbative expansion in $J$, since the $\n$ propagators also include a self-consistent renormalization from the disorder averaging.  

The ladder series in Fig.\ref{ladsum}(a) can be evaluated in a straightforward fashion and is given by a geometric series when expressed in terms of,
\beq
F(\omega_1,\omega_2)\equiv\int\int F(t_1,t_2) ~e^{i (\omega_1 t_1+\omega_2 t_2)}~ dt_1~ dt_2.
\eeq 
The sum evaluates to,
\begin{equation}
F(\omega_1,\omega_2)=\frac{N}{M}\frac{Q_R(\omega_1)Q_R(\omega_2)}{1-J^2Q_R(\omega_1)Q_R(\omega_2)}.\label{FF}
\end{equation}

The exponential growth in time, if any, of the squared commutators in this case will arise from the (imaginary) poles of $F(\omega_1,\omega_2)$ in Eq. \ref{FF}. However a simple exercise immediately shows that there are no such poles of $F(\omega_1,\omega_2)$ and as a result $F(t_1,t_2)$ does {\it not} exhibit any exponential growth with $\lambda_L>0$.

\begin{figure}[h]
\begin{center}
    \includegraphics[scale=0.25]{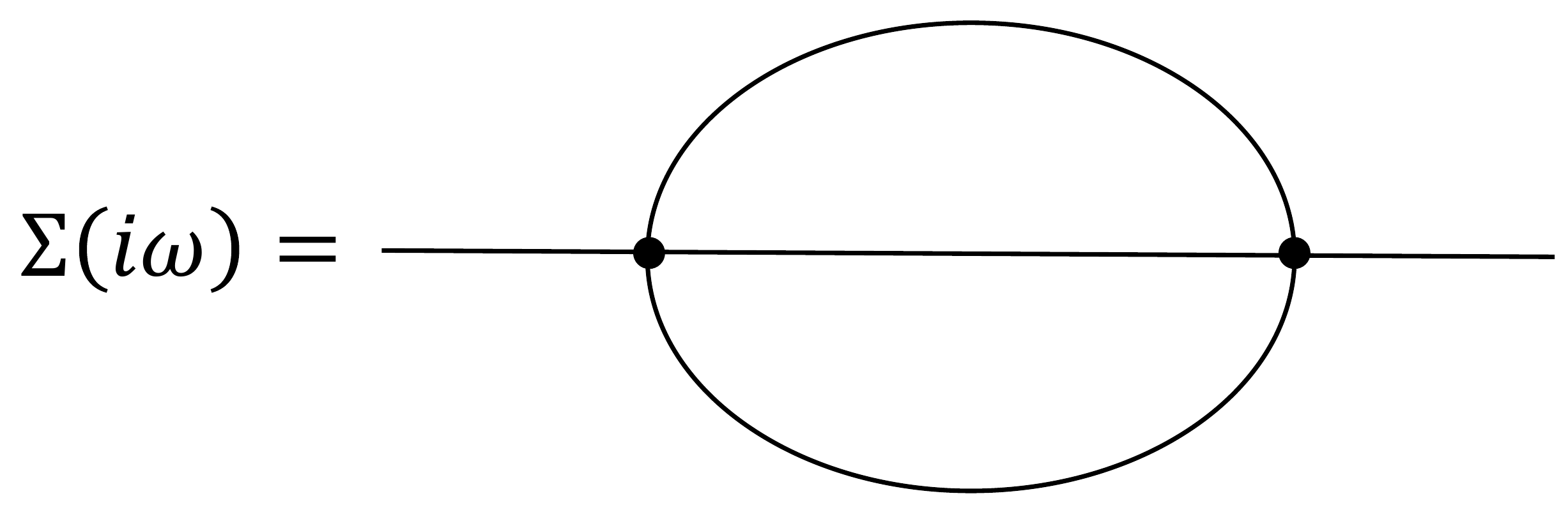}
\end{center}
\caption{The leading non-analytic contribution to $\Sigma(i\omega)$ at $O(u^2/M)$ (Eq.~\ref{se}).}
\label{SEu}
\end{figure}

{\it Corrections beyond large $M$.-} In order to obtain the first non-trivial contribution to $\lambda_L$, we have to consider corrections beyond the $M\rightarrow\infty$ limit considered thus far. Previous work \cite{SS1,SS2} has shown that the $1/M$ corrections in the disorder-averaged theory can be studied systematically within a soft-spin formulation of the problem. The leading $1/M$ correction arises from the four-body interaction term $\sim u(\n^2)^2$, which modifies the ladder series for $F(t_1,t_2)$ (see Fig.\ref{ladsum}(b)) and introduces a correction to the self-energy for $Q_R(\omega)$ (Fig.\ref{SEu}). As has been noted earlier \cite{DSweak,BSDC17,PatelSachdev,DCBS17}, the former can contribute to an exponential growth and the latter to an exponential decay for $F(t_1,t_2)$, which may lead to a net exponential growth  with $\lambda_L>0$.

Including the $1/M$ correction, the propagator has the form

\beq
Q(i\omega_n)&=&\frac{g}{\omega_n^2+\lambda-gJ^2 Q(i\omega_n)-\frac{\Sigma(i\omega_n)}{M}},
\label{1M}
\eeq

which can be solved in a straightforward fashion and where the leading non-analytic correction to the self-energy (Fig.\ref{SEu}), after analytically continuing to real frequencies \cite{si} gives
\beq
\Sigma_R(\omega)\sim u^2 |\omega|^5~\tn{sgn}(\omega),
\label{se}
\eeq
which is less singular than the $M\rightarrow\infty$ saddle point correction and hence does not affect the long time behavior of $Q(t)$. However, the presence of this self-energy correction immediately leads to an exponential decay for $F(t_1,t_2)\sim e^{-\Gamma(t_1+t_2)}$ with a small decay rate $\Gamma=\alpha(u^2/M) \Delta_s^4(T)$, where $\alpha$ is a known function of $g,J$ \cite{si}.

\begin{figure}[h]
\begin{center}
    \includegraphics[scale=0.45]{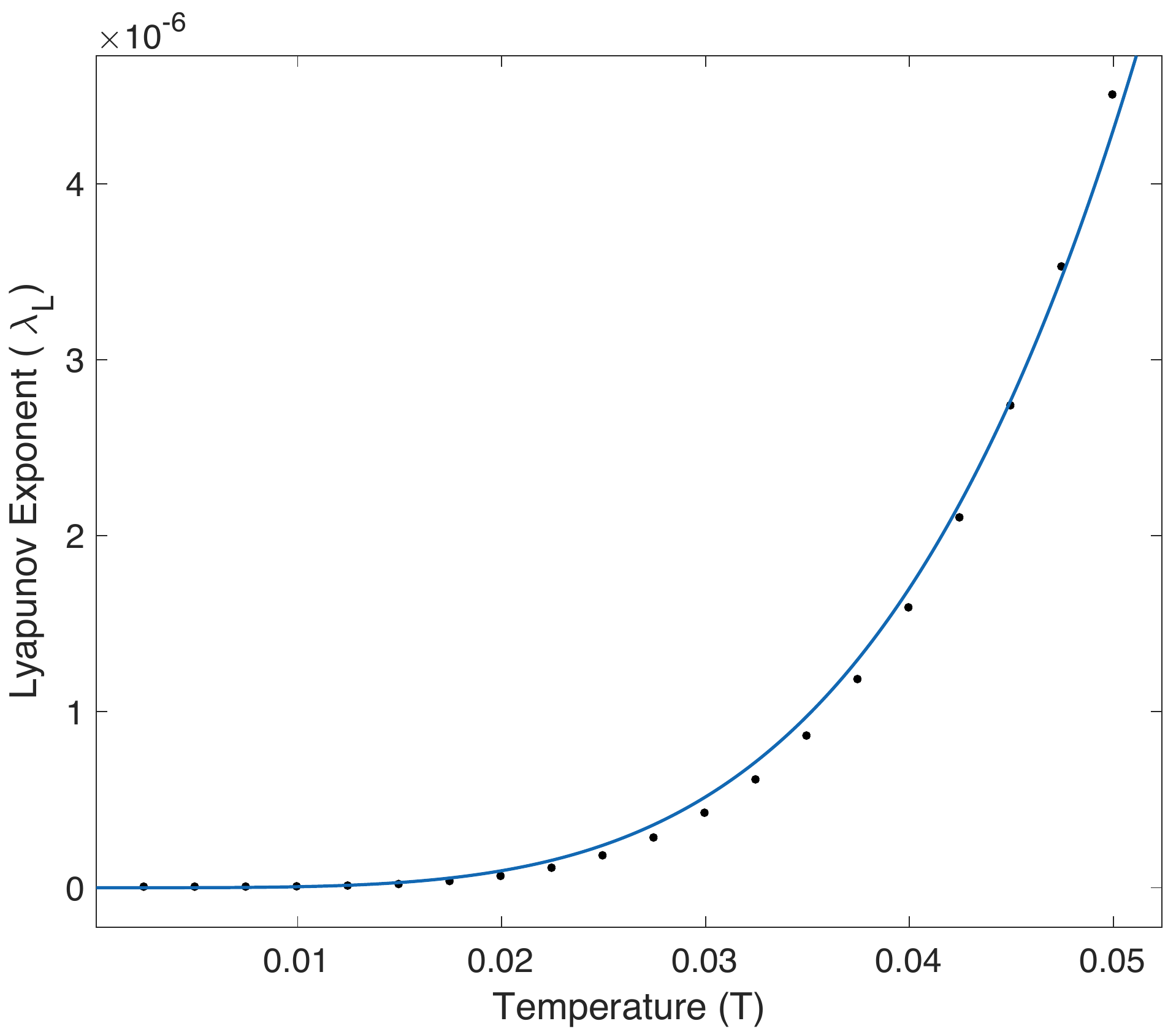}
\end{center}
\caption{Numerically evaluated growth exponent, $\lambda_L\approx\Upsilon$, for $F(t,t)$ at low temperature. Black markers denote numerically evaluated datapoints and solid line denotes a fit to the functional form $\Upsilon\sim T^3\Delta_s(T)$. Temperature is measured relative to a UV cutoff of $\Lambda ~(=J) =1$ and $\lambda_L$ is measured in units of inverse time with $g = 9\pi^2/16 J,~ u^2/M =0.01 J^2$.}
\label{expo}
\end{figure}

On the other hand, the ladder series for $F(t_1,t_2)$ itself is modified due to the addition of a new `rung' (Fig.\ref{ladsum} (b)), such that the self-consistent Bethe-Salpeter equation is given by
\beq
F(t_1,t_2)&=&Q_R(t_1)Q_R(t_2)+J^2\int_{t_3,t_4}~Q_R(t_{13})~Q_R(t_{24})~F(t_3,t_4)\nonumber\\
&+&\frac{2u^2}{M}\int_{t_3,t_4}~Q_R(t_{13})~Q_R(t_{24})~Q_W^2(t_{34})~F(t_3,t_4),
\eeq
where 
\beq
Q_W(t)~\delta_{ij} \delta_{ab} &=& \tn{Tr}\{\sqrt\rho~ n_i^a(t)\sqrt{\rho}~ n_j^b(0) \}
\eeq
is the Wightman Green's function. Note that the first line in the expression for $F(t_1,t_2)$ represents the contribution that we already evaluated earlier while the second line represents the contribution from the new rung. The exponential growth of $F(t_1,t_2)$, if any, will arise from the eigenvalues for the second line above, i.e. $\lambda_L\sim O(u^2/M)$. At weak coupling (i.e. small $u$), we are ignoring here another contribution to the ladder series at $O(u^4/M)$. In the $\omega\ll\sqrt{gJ}$ limit,
\begin{equation}
Q_W(\omega) = \frac{A(\omega)}{2\sinh(\beta\omega/2)} \approx
\begin{cases}
\frac{\sqrt{gJ}}{gJ^2}\frac{\tn{sgn}(\omega)\sqrt{\omega^2-\Delta_s^2}}{2\sinh(\beta\omega/2)},~~|\omega|>\Delta_s\\
0, ~~~~~~~~~~~~\tn{otherwise}.
\end{cases}
\end{equation}
(See Appendix C of Ref.~\cite{DCBS17}). An explicit numerical evaluation of the contribution to the small exponentially growing piece leads to $F(t_1,t_2)\sim  e^{\Upsilon(t_1+t_2)}$ with $\Upsilon = \eta(u^2/M)T^3\Delta_s(T)$, where $\eta$ is a known function of $g,~J$ \cite{si}. Thus, at low temperatures $\Upsilon$ is parametrically larger than $\Gamma$ with  $\lambda_L=(\Upsilon - \Gamma)>0$, resulting in an overall exponential growth of $F(t_1,t_2)$. Thus at low temperatures in the quantum critical regime, we find $\lambda_L \sim (u^2/M) [T^4/\log^{1/2}(1/T)]$ as shown in Fig. \ref{expo}.

{\it Integrability and random matrix model.-}   
Let us now consider a collection of $M$ component vector fields, $\vec{V}_p$, with a unit constraint $\vec{V}_p^2=1$, where $p=1,..,N$. Let the `frequencies' (or `mass') terms for these fields be drawn from the distribution of eigenvalues of a large $N\times N$ random matrix $J_{\ell m}$, whose elements are independently distributed, such that $\sum_m J_{\ell m}\vp_m^{(p)}=\lambda_{p}\vp_\ell^{(p)}$. 
The $\lambda_p$ are the eigenvalues and $\vp^{(p)}$ denote the eigenvectors of the random matrix $J_{\ell m}$. For any given realization of $J_{\ell m}$, the Lagrangian can be written as,
\beq
\mathcal{S}=-\int d\tau \sum_{p=1}^N\frac{M}{2}\bigg[\frac{1}{g}[(\partial_\tau \vec{V}_p)^2+\lambda(\vec{V}_p)^2]+ \frac{1}{\sqrt{N}}\lambda_p(\vec{V}_p)^2 +  ...\bigg],
\label{lag_M}
\eeq
where $\lambda$ is a Lagrangian multiplier that imposes the unit constraint and the action uses a specific normalization with respect to $M,~N$ in order to relate to the saddle-point action of the random rotor model. The ellipses denote additional terms beyond quadratic order that we leave unspecified for now. 

The auto-correlator for the field $\vec{V}_p$ in any given realization is given by,
\beq
\widetilde{Q}_p(i\omega_n) = \frac{1}{(\omega_n^2+\lambda)/g+\lambda_p/\sqrt{N}}.
\eeq
It is then natural to ask how the disorder averaged auto-correlator behaves as a function of frequency, after averaging over the distribution of different $\lambda_p$. For the Gaussian orthogonal ensemble (GOE) of random matrices, the level density (or density of states), $\sigma^{}_J(\lambda_p)$, is given by,
\beq
\sigma^{}_J(x)=\frac{1}{2\pi J}\sqrt{4N-(x/J)^2},
\eeq
otherwise also known as the `Wigner semi-circle law' \cite{mehta}. The disorder averaged auto-correlator $Q(i\omega_n)$ can be evaluated,
\beq
Q(i\omega_n) &=& \frac{1}{N}\int_{-2J\sqrt{N}}^{2J\sqrt{N}} d\lambda_p~ \widetilde{Q}_p(i\omega_n)~\sigma^{}_J(\lambda_p),\nonumber\\
&=& \frac{1}{J}\left[\frac{\omega_n^2+\lambda}{2gJ}-\sqrt{\left(\frac{\omega_n^2+\lambda}{2gJ}\right)^2-1}\right].
\eeq
It is not a coincidence that the above propagator is identical to the propagator for the rotor-fields, obtained upon solving Eqn.~\ref{SC}.

The mapping between the two seemingly unrelated models can be made precise by considering the following orthogonal transformation
\begin{equation}
\vec{n}_i(\tau)=\sum_{p=1}^N \vp^{(p)}_i \vec{V}_p(\tau).
\end{equation}
The problem is thus essentially free in the $N,~M\rightarrow\infty$ limit, described by a random-matrix, and is responsible for the absence of an exponential growth of $F(t_1,t_2)$. We can similarly express the quartic interaction between the rotor fields in terms of the fields introduced above and systematically compute the leading $1/M$ corrections to the self-energy \cite{si}. Disorder averaging now requires a knowledge of the distributions of both the eigenvalues, $\lambda_p$ and the eigenvectors, $\vp_i^p$, which are known to be independent for GOE random-matrices \cite{RMTbook}. Within the random-matrix formulation of the problem, the correlations among the different eigenvalues are also included in the $n-$level ``cluster functions" \cite{RMTbook}. The replica-action for the rotor problem does not include these correlations and assumes that the eigenvalues are completely independent; the completely uncorrelated piece in the cluster function, which factorizes into a product of the $\sigma_J(\lambda_i)$, reproduces the results obtained using the replica action \cite{si}. The alternative formulation in terms of the random-matrix therefore allows us to clearly disentangle the features that arise from disorder averaging and interactions.

{\it A generalized model.-} Inspired by the structure of the saddle-point action of the $O(M)$ random rotor model, we propose here an extension of the model to instead include an interaction between $p$-pairs of rotors (instead of $p=1$)
\begin{equation}
    H=g\sum_i \frac{\hat{L_i}^2}{2M}+\frac{M^{\frac{p+1}{2}}}{\sqrt{p}N^{p-1/2}}\sum_{\{i,j\}}J_{i_1j_1...i_pj_p}\prod_{k=1}^p(\vec{n}_{i_k}\cdot\vec{n}_{j_k}),
\end{equation}
where $J_{i_1j_1...i_pj_p}$, symmetric under $i_k\leftrightarrow j_k$ and $(i_k,j_k)\leftrightarrow(i_l,j_l)$, are Gaussian random variables with mean zero and finite variance, $J^2$. After integrating over disorder and assuming a replica-symmetric solution, the saddle-point equations become,
\beq
Q(i\omega_n) &=& \frac{g}{\omega_n^2 + \lambda - \Sigma(i\omega_n)},\\
\Sigma(\tau) &=& gJ^2 [Q(\tau)]^{p}.
\eeq
 However the scaling behavior at the QCP between a paramagnet and a spin-glass, {\it if} it exists {\footnote{The above model while different from the quantum spherical $p-$spin glass model, bears some superficial resemblance at the level of the replica action. It has been argued \cite{LC01} that for the latter, there is no continuous QPT between the paramagnet and spin-glass for $p>2$, and the transition is first-order ($p=2$ reduces to the usual random rotor model \cite{SS1} considered earlier).}}, can be obtained using scaling (when $\Sigma(i\omega=0)=\lambda$). For $p>1$, this yields,
\beq
Q(\tau) &\sim& \frac{\tn{sgn}(\tau)}{\tau^{2\Delta(p)}},\\
\Sigma(i\omega) &\sim& \tn{sgn}(\omega) |\omega|^{1-2\Delta(p)},
\eeq
where $\Delta(p)=1/(p+1)$. 

At this point, the correspondence between the generalized rotor models with $p\geq1$ and the SYK$_q$ models \cite{sykq} should be clear. The QCP for the rotor model with $p=1$ bears striking similarity with SYK$_2$, a random-matrix model, and explains the lack of any exponential growth of the OTOC. On the other hand, the QCP for the rotor models with $p>1$ have a scaling structure that is similar to the SYK$_q$ models. We leave a detailed analysis of our model for future work.

{\it Discussion.-} In this work, we have studied the OTOC for a rotor model interacting with random exchange interactions at the QCP between a paramagnet and a spin-glass. There are interesting connections between the large$-N,~M$ saddle point action and a purely random-matrix model supplemented with $O(M)$ invariant quartic interactions. It will be interesting to study the leading $1/N$ corrections to the saddle point solution using the properties of the asymptotic Tracy-Widom probability distribution \cite{SNM} of the largest eigenvalues of a random matrix.

{\it Acknowledgements.-} DC is supported by a postdoctoral fellowship from the Gordon and Betty Moore Foundation, under the EPiQS initiative, Grant GBMF-4303 at MIT. TS is supported by a US Department of Energy grant DE-SC0008739, and in part by a Simons Investigator award from the Simons Foundation.

{\it Note added.-} While this manuscript was being finalized for submission, we became aware of related results \cite{GCBS}.

\bibliographystyle{apsrev4-1_custom}
\bibliography{scrambling}
\begin{widetext}
\section{SUPPLEMENTARY MATERIAL for ``Slow scrambling and hidden integrability in a random rotor model"}

\section{Replica action for random $O(M)$ rotor model}
\label{saddle1}
In this subsection, we provide some additional details for the large $N,~M$ saddle point treatment for the rotor model. To begin, recall that the commutation relations for the rotor fields and the angular momenta are given by,
\beq
[L_{i\mu\nu},n_{j\sigma}] = i\delta_{ij} (\delta_{\mu\sigma}n_{j\nu} - \delta_{\nu\sigma}n_{j\mu}).
\eeq
In the $N\rightarrow \infty$ limit, the saddle point action can be obtained upon integrating over disorder configurations. The replicated partition function in Euclidean time is then given by,
\begin{eqnarray*}
Z_n&=&\int \mathcal{D}Q^{aa}_{mm}(\tau,\tau')~\mathcal{D}Q^{aa}_{mn}(\tau,\tau')~\mathcal{D}P^{ab}_{mn}(\tau,\tau')~\mathcal{D}\lambda~\mathcal{D}n^a(\tau)\\
&&\tn{exp}\bigg\{\int_0^\beta d\tau \int_0^\beta d\tau'\frac{NJ^2}{2}\bigg[-\sum_{a,m}\frac{1}{2}(Q^{aa}_{mm})^2-\sum_{a,m<n}(Q^{aa}_{mn})^2-\sum_{a<b,m,n}(P^{ab}_{mn})^2\bigg]\\
&&+\frac{MJ^2}{2}\bigg[\sum_{a,m}Q^{aa}_{mm}(\tau,\tau') \sum_i n_i^{ma}(\tau)n_i^{ma}(\tau')+2\sum_{a,m<n}Q^{aa}_{mn} \sum_i n_i^{ma}n_i^{na}+2\sum_{a<b,m,n}P^{ab}_{mn}\sum_i n_i^{ma}n_i^{nb}\bigg]\\
&&-\frac{M}{2g}\int_0^\beta d\tau\sum_{i,a}[(\partial_{\tau}n_i^a(\tau))^2+\lambda n_i^a(\tau))^2]\bigg\}
\end{eqnarray*}
where $a,b\in\{1,...,n\}$ are replica indices, $i,j\in\{1,...,N\}$ are site indices, and $m,n\in\{1,...,M\}$ are vector indices. $Q_{mn}^{aa}$ and $P_{mn}^{ab}$ correspond to quadrupolar and spin glass order, respectively. Since we are only interested in the paramagnetic regimes in this work, both of the above quantities have zero expectation value.

Assuming the saddle point to be $O(M)$ invariant, the replica action can be simplified such that it describes the quantum mechanical action for a single rotor with multiple replica indices,
\beq
Z=\int \mathcal{D}n^a(\tau)~\mathcal{D}\lambda~\mathcal{D}Q^{ab}~ \tn{exp}\bigg[-\frac{M}{2g}\int d\tau[(\partial_\tau n^a)^2+\lambda((n^a)^2-1)]\nonumber\\
+\frac{MJ^2}{2}\int d\tau\int d\tau'[Q^{ab}(\tau-\tau')n^a(\tau)\cdot n^b(\tau')-\frac{1}{2}Q^{ab}(\tau-\tau')^2]\bigg].
\eeq
In the large $M$ limit, we can define the imaginary time ($\tau$) auto-correlation functions $Q(\tau)=\langle \n(\tau)\cdot \n(0)\rangle$, where we are only interested in regimes where $Q(\tau)$ is replica diagonal. We can solve for $Q(i\omega_n)$ (which leads to the above self-consistency condition Eq. \ref{SC}) and obtain,
\begin{equation}
Q(i\omega_n)=\frac{\omega_n^2}{2gJ^2}+\frac{\lambda}{2gJ^2}-\frac{1}{2gJ^2}\sqrt{(\omega_n^2+\lambda-2gJ)(\omega_n^2+\lambda+2gJ)}.
\label{q}
\end{equation}

At $T=0$, the spin-gap vanishes at the critical point between a paramagnet and a spin-glass phase when $\lambda=2Jg$ (see fig.\ref{phasediag}); the actual phase-boundary in the $g-T$ plane can be obtained by solving for the constraint equation $Q(\tau=0)=1$ with the above value of $\lambda$, which can be recast as,
\beq
\int_0^\infty \frac{d\omega}{\pi} A(\omega) \coth\bigg(\frac{\beta\omega}{2} \bigg) = 1.
\label{cons}
\eeq
Near the $T=0$ quantum critical point, this is given by,
\beq
gJ = \frac{9\pi^2J^2}{16} - 3T^2.
\eeq

\begin{figure}[h!]
\begin{center}
\includegraphics[scale=0.25]{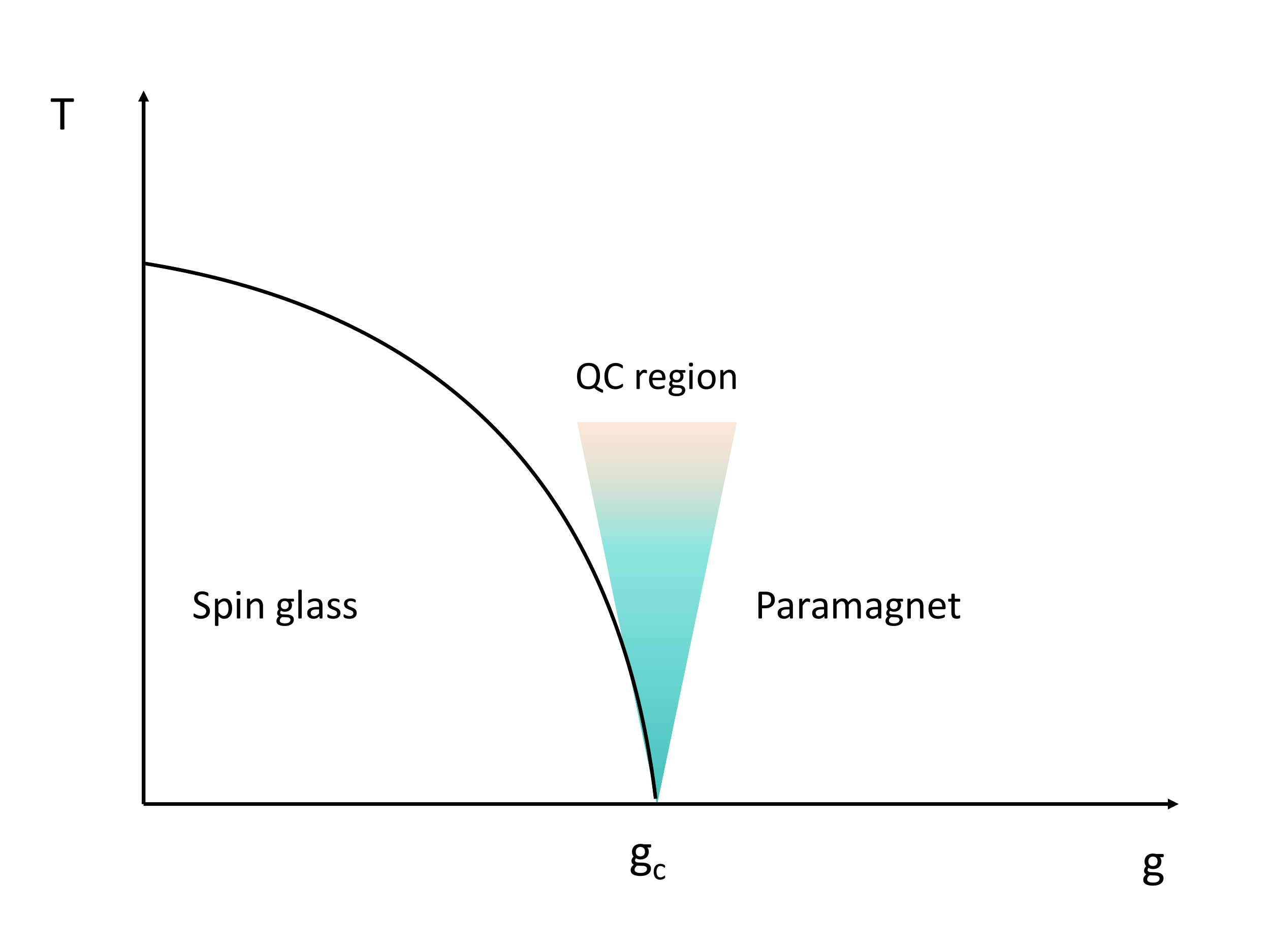}
\end{center}
\caption{The phase-diagram for the random rotor model as a function of temperature and $g$.}
\label{phasediag}
\end{figure}

\section{$F(t_1,t_2)$ in the $N,~M\rightarrow\infty$ limit}
\label{ft}

In this subsection, we obtain the time dependence of $F(t_1,t_2)$ in the large $N,~M$ limit, where there is no exponential growth. Recall that,
\begin{equation}
F(\omega_1,\omega_2)=\frac{N}{M}\frac{Q_R(\omega_1)Q_R(\omega_2)}{1-J^2Q_R(\omega_1)Q_R(\omega_2)}=\frac{N}{MJ^2}\bigg(-1+\frac{1}{1-J^2Q_R(\omega_1)Q_R(\omega_2)}\bigg).
\end{equation}
The quantum critical region of interest to us here is when $\Delta_s^2\ll gJ$. Let us introduce the variables,
\beq
x=\sqrt{\frac{-\omega_1^2+\Delta_s^2}{2gJ}},~~y=\sqrt{\frac{-\omega_2^2+\Delta_s^2}{2gJ}},
\eeq
such that the denominator can be simplified to,
\begin{eqnarray}
1-J^2Q_R(\omega_1)Q_R(\omega_2)&\approx&1-(x^2+1-x\sqrt{2})(y^2+1-y\sqrt{2})\\
&=& - (x+y)(x+y-\sqrt{2}),
\end{eqnarray}
where we have assumed $x,y\ll1$ in the first line above and ignored terms that are $O(xy^2,~x^2y)$ and higher. 
By making this approximation, we only focus on the long time, long period behavior with a characteristic time scale $1/\sqrt{2gJ}$.

We can then approximate $F(x,y)$ as
\begin{eqnarray}
F(x,y)&\approx&\frac{N}{MJ^2}\bigg[-1+\frac{1}{\sqrt{2}}\bigg(\frac{1}{x+y}+\frac{1}{\sqrt{2}-x-y}\bigg)\bigg]\\
&\approx&\frac{N}{MJ^2}\bigg[-\frac{1}{2}+\frac{1}{\sqrt{2}}\frac{1}{x+y}+O(x,y)\bigg].
\end{eqnarray}
The leading time dependence of $F(t_1,t_2)$ comes from the  second term above, which can be simplified as,
\begin{eqnarray}
\frac{1}{x+y}&=&\frac{\sqrt{2gJ}}{\Delta_s}\bigg[\frac{\sqrt{1-(\omega_1^2/\Delta_s^2)}-\sqrt{1-(\omega_2^2/\Delta_s^2)}}{-(\omega_1^2/\Delta_s^2)+(\omega_2^2/\Delta_s^2)}\bigg].
\label{1byxy}
\end{eqnarray}
Rescaling frequencies and time as $\tilde\omega_{1,2}=\omega_{1,2}/\Delta_s,~ \tilde{t}_{1,2}=\Delta_s t_{1,2}$, the inverse Fourier transform of Eqn.~\ref{1byxy} is given by

\begin{eqnarray}
I(\tilde{t}_1,\tilde{t}_2)\equiv\sqrt{2gJ}\Delta_s\int_{-\infty}^{\infty}\int_{-\infty}^{\infty}\frac{\sqrt{1-\tilde\omega_1^2}-\sqrt{1-\tilde\omega_2^2}}{(\tilde\omega_2+\tilde\omega_1)(\tilde\omega_2-\tilde\omega_1)}e^{-i\tilde\omega_1\tilde{t}_1-i\tilde\omega_2\tilde{t}_2}\frac{d\tilde\omega_1}{2\pi}\frac{d\tilde\omega_2}{2\pi}
\label{int_I}
\end{eqnarray}
We can redefine $\tilde{t} = \frac{\tilde{t}_1-\tilde{t}_2}{2}$ and $\tilde{T} = \frac{\tilde{t}_1+\tilde{t}_2}{2}$. $I(\tilde{t}_1,\tilde{t}_2) = I(\tilde{T},\tilde{t})$. Taking partial derivative over Eqn.$\tilde{t}$ in \ref{int_I} yields,
\begin{equation}
    \frac{\partial I(\tilde{T},\tilde{t})}{\partial \tilde{t}} = i  \sqrt{2gJ}\Delta_s\int_{-\infty}^{\infty}\int_{-\infty}^{\infty}\frac{\sqrt{1-\tilde\omega_1^2}-\sqrt{1-\tilde\omega_2^2}}{\tilde\omega_2+\tilde\omega_1}e^{-i\tilde\omega_1\tilde{t}_1-i\tilde\omega_2\tilde{t}_2}\frac{d\tilde\omega_1}{2\pi}\frac{d\tilde\omega_2}{2\pi}\label{deriv_I_t}
\end{equation}
We can integrate over $\tilde{\omega}_2$ in the first term in Eqn.~\ref{deriv_I_t}, and integrate over $\tilde{\omega}_1$ in the second term in Eqn.~\ref{deriv_I_t}.
\beq
\frac{\partial I(\tilde{T},\tilde{t})}{\partial \tilde{t}}=  \frac{\sqrt{2gJ}\Delta_s}{2} \Big(sgn(\tilde{t}_2)\int_{-\infty}^{\infty}\sqrt{1-\tilde\omega_1^2}e^{-2i\tilde\omega_1\tilde{t}}\frac{d\tilde\omega_1}{2\pi}-sgn(\tilde{t}_1)\int_{-\infty}^{\infty}\sqrt{1-\tilde\omega_2^2}e^{2i\tilde\omega_2\tilde{t}}\frac{d\tilde\omega_2}{2\pi}\Big)=\sqrt{2gJ}\Delta_s sgn(\tilde{t}) \frac{J_1(2\tilde{t})}{4\tilde{t}},\label{partial_t_tilde}
\eeq
where $J_1$ is the Bessel function and we always let $\tilde{t}_{1,2} \geq 0$.

Let me explain in more detail how to perform the square-root integral,
\beq
\int_{-\infty}^{\infty} \sqrt{1-\omega^2} e^{-i \omega t} \frac{d\omega}{2\pi} &=& \int_{-1}^{1} \sqrt{1-\omega^2} e^{-i \omega t} \frac{d\omega}{2\pi} -i\int_{1}^{\infty} \sqrt{\omega^2-1} e^{-i \omega t} \frac{d\omega}{2\pi}+i\int_{-\infty}^{-1} \sqrt{\omega^2-1} e^{-i \omega t} \frac{d\omega}{2\pi}\\
&=& \frac{J_1(t)}{2t} - \frac{K_1(i t)+K_1(-i t)}{2 \pi t}\\
&=& \theta(t)\frac{J_1(t)}{t},\label{bessel}
\eeq
where $\theta(t)$ is the Heaviside function and $K_1$ is the Bessel function of the second kind.

Using Eqn.\ref{bessel}, we can get the result in Eqn.\ref{partial_t_tilde}. Integrate over $\tilde{t}$ in Eqn.\ref{partial_t_tilde}, we can get,
\beq
I(\tilde{T},\tilde{t}) = I(\tilde{T},0) + \frac{\sqrt{2gJ}\Delta_s}{4} |\tilde{t}| [_1F_2](\frac{1}{2};\frac{3}{2},2;-\tilde{t}^2),
\eeq
where $[_1F_2]$ is the hypergeometric function.

Next, we want to calculate $I(\tilde{T},0)$.
\beq
\frac{d I(\tilde{T},0)}{d \tilde{T}} &=& -i \sqrt{2gJ}\Delta_s \int_{-\infty}^{\infty}\int_{-\infty}^{\infty}\frac{\sqrt{1-\tilde\omega_1^2}-\sqrt{1-\tilde\omega_2^2}}{\tilde\omega_2-\tilde\omega_1}e^{-i(\tilde\omega_1+\tilde\omega_2)\tilde{T}}\frac{d\tilde\omega_1}{2\pi}\frac{d\tilde\omega_2}{2\pi}\\
&=&-i 2\sqrt{2gJ}\Delta_s \int_{-\infty}^{\infty}\int_{-\infty}^{\infty}\frac{\sqrt{1-\tilde\omega_1^2}}{\tilde\omega_2-\tilde\omega_1}e^{-i(\tilde\omega_1+\tilde\omega_2)\tilde{T}}\frac{d\tilde\omega_1}{2\pi}\frac{d\tilde\omega_2}{2\pi}\\
&=&-\sqrt{2gJ}\Delta_s sgn(\tilde{T})\int_{-\infty}^{\infty}\sqrt{1-\tilde\omega_1^2}e^{-2i\tilde\omega_1\tilde{T}}\frac{d\tilde\omega_1}{2\pi}\\
&=&-\sqrt{2gJ}\Delta_s sgn(\tilde{T})\frac{J_1(2\tilde{T})}{2\tilde{T}}
\eeq
Integrate over $\tilde{T}$, one can get,
\beq
I(\tilde{T},0) &=& I(\infty,0) - \frac{\sqrt{2gJ}\Delta_s}{2} \tilde{T} [_1F_2](\frac{1}{2};\frac{3}{2},2;-\frac{(2\tilde{T})^2}{4})+\frac{\sqrt{2gJ}\Delta_s}{2}\\
&=&-\frac{\sqrt{2gJ}\Delta_s}{4} (\tilde{t}_1+\tilde{t}_2) [_1F_2](\frac{1}{2};\frac{3}{2},2;-\frac{(\tilde{t}_1+\tilde{t}_2)^2}{4})+\frac{\sqrt{2gJ}\Delta_s}{2}
\eeq
Note that $\lim_{t_1\to \infty, t_2\to\infty}F(t_1,t_2) =0$ by Riemann-Lebesgue lemma. Here, we approximate $I(\tilde{T},0)\approx 0$ in order to match the long time behavior of $F(t_1,t_2)$.

Finally, we obtain the approximate expression for the squared-commutator,
\beq
F(t_1,t_2) \approx \frac{N\sqrt{gJ}\Delta_s}{MJ^2}\Big[\frac{1}{2}-\frac{1}{4}(\tilde{t}_1+\tilde{t}_2) [_1F_2](\frac{1}{2};\frac{3}{2},2;-(\tilde{t}_1+\tilde{t}_2)^2/4)+\frac{1}{8}|\tilde{t}_1-\tilde{t}_2| [_1F_2](\frac{1}{2};\frac{3}{2},2;-(\tilde{t}_1-\tilde{t}_2)^2/4)\Big].
\eeq

\section{Self-energy at $O(1/M)$ and decay rate for OTOC}
\label{SEsup}
As discussed in the main text, the leading correction to the self-energy at $O(u^2/M)$ from the quartic interaction term (Fig.~\ref{SEu}) is given by,
\beq
\frac{\Sigma(i\omega_n)}{g} &=& u^2\left(\frac{2\sqrt{gJ}}{gJ^2}\right)^3\int d\tau~ \frac{1}{\tau^6}~e^{-i\omega_n \tau}
\nonumber\\
&=&-u^2\left(\frac{\sqrt{gJ}}{gJ^2}\right)^3\frac{1}{15}\pi |\omega_n|^5.
\label{se}
\eeq

The above self-energy correction gives rise to an exponential decay of the Green's function and to the OTOC. The decay rate can be estimated in a straightforward fashion by looking for the imaginary solutions for the following equation,
\begin{equation}
-\omega^2+\Delta_s^2-\frac{\Sigma(\omega)}{M}=0.
\end{equation}
This yields, $\omega=\pm\Delta_s-i\Gamma$, where
\beq
\Gamma= \frac{u^2}{M}\frac{g}{30}\bigg(\frac{\sqrt{gJ}}{gJ^2}\bigg)^3\Delta_s^4,
\eeq
as quoted in the main text. The negative imaginary part gives rise to an exponential decay of $F(t_1,t_2)$.

\section{Numerical analysis of $1/M$ corrections}
In this subsection, we describe details of our numerical evaluation of the ladder sum for $F(t,t)$. The Bethe-Salpeter equation for the $1/M$ corrected squared-commutator $F_u(t_1,t_2)$ can be written as,
\begin{equation}
F_u(t_1,t_2)=F_d(t_1,t_2)+\frac{2u^2}{M}\int_0^{t_1} dt_3 \int_0^{t_2} dt_4~ F_d(t_1-t_3,t_2-t_4)~P(t_3-t_4)~F_u(t_3,t_4),\label{int_num}
\end{equation}
where $F_d(t_1,t_2)$ represents the contribution to the ladder sum without the $O(u^2/M)$ correction to the rung, but includes the self-energy at $O(u^2/M)$ in the dressed propagators. As noted earlier, $F(t_1,t_2)$ in the large $N,~M$ limit does not have any exponential growth; an explicit analytical form for $F(t_1,t_2)$ appears below. Upon including the $O(u^2/M)$ self-energy correction into account, $F_d(t_1,t_2) =F(t_1,t_2)e^{-\Gamma (t_1+t_2)}$,  where we evaluated $\Gamma \sim (u^2/M) \Delta_s^4$. Finally, we have defined $P(t)=[Q_W(t)]^2$.

The above self-consistent integral equation can be viewed as solving a matrix inversion problem once we rewrite it as,
\begin{equation}
\int dt_3 \int dt_4~ \bigg[\delta(t_{13})~\delta(t_{24})-\frac{2u^2}{M} ~F_d(t_{13},t_{24})~P(t_3-t_4)\bigg]~F_u(t_3,t_4)=F_d(t_1,t_2).
\end{equation}
In more explicit terms, the integral on the left hand side can be discretized as a summation on a fine (time) grid,  $[A_{t_1t_2;t_3t_4} F_{u;t_3t_4}] = F_{d;t_1t_2}$, such that computing the inverse $A_{t_1t_2;t_3t_4}^{-1}$ will lead us to the required form of $F_u$.

In our numerical calculations, we focus exclusively on the exponential piece of $F_u(t,t)=F_0(t)~e^{\lambda_L t}$, where $F_0(t)$ is an undetermined function of time and $\lambda_L>0$ is the Lyapunov exponent. At late times, our numerical analysis is consistent with an exponent,  $\lambda_L=a\frac{T^4}{\sqrt{\log(1/T)}}$, where $a$ is a temperature independent `fitting' parameters. However, there is some uncertainty associated with observing the $1/\sqrt{\log(1/T)}$ piece in $\lambda_L$ in our numerical analysis. 

The numerical result for $F_u(t,t)$ is summarized in Fig.\ref{num1} for different temperature.

\begin{figure}[h!]
\begin{center}
\includegraphics[scale=.6]{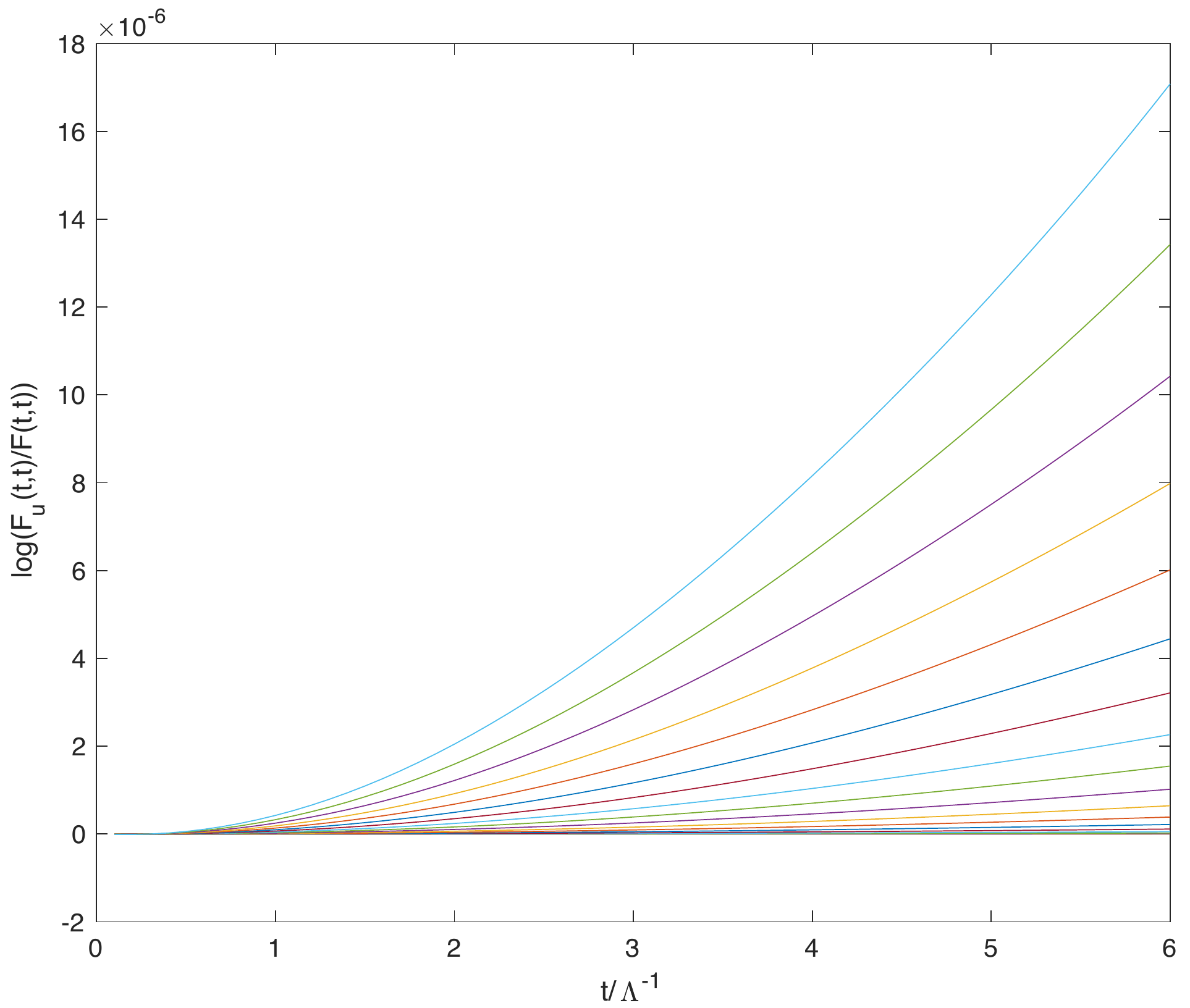}
\end{center}
\caption{A plot of $\log[F_u(t,t)/F(t,t)]$ as a function of time for different temperatures shows a clear exponential growth. The lines from top to bottom represent temperature from $0.05\Lambda$ to $0.025\Lambda$ with interval $0.025\Lambda$.}
\label{num1}
\end{figure}

\section{$1/M$ corrections to the random matrix model}
The quartic interaction between the rotor fields can be written as,
\beq
H_{\tn{int}} = uM \sum_i \bigg[\vec{n}_i^2(\tau)\bigg]^2= uM \sum_i \sum_{\{p_i\}} \vp_i^{p_1}\vp_i^{p_2}\vp_i^{p_3}\vp_i^{p_4} (\vec{V}_{p1}\cdot\vec{V}_{p2})(\vec{V}_{p3}\cdot\vec{V}_{p4}),
\eeq
where $\vp_i^p$ represents an orthogonal matrix. 

Let us consider the $1/M$ correction to the Green's function for the $\vec{V}_p$ field due to the above interaction term. At one-loop order, this leads to a contribution $O(u/M)$ which renormalizes the `mass' with no frequency dependence and can be absorbed into a redefinition of $\lambda$. At two-loop order, the contribution of the interaction to the self-energy is,
\beq
\Sigma_{p_1p_1'} = u^2  \sum_{i,j}\sum_{p_2p_3p_4} \vp_i^{p_1}\vp_i^{p_2}\vp_i^{p_3}\vp_i^{p_4}\vp_j^{p_1'}\vp_j^{p_2}\vp_j^{p_3}\vp_j^{p_4}~ \widetilde{Q}(\lambda_{p_2})~\widetilde{Q}(\lambda_{p_3})~\widetilde{Q}(\lambda_{p_4})
\label{sermt}
\eeq
where $\widetilde{Q}(\lambda_p)$ is the Green's function for $\vec{V}_p$, as introduced earlier, and we have suppressed the frequency dependence above.

In order to carry out an averaging over the disorder distribution, we now need to know the distribution of both the eigenvalues, $\lambda_p$ and the eigenvectors, $\vp_i^p$; for the GOE of random-matrices, these distributions are known to be {\it independent} \cite{RMTbook}. This allows us to integrate over the eigenvalues first. (Note that the bare Green's function for one component of $\vec{V}_p$ has a factor $1/M$ such that the leading $1/M$ correction has a factor $1/M^2$.) Including the outer legs, we have,
\beq
\frac{1}{M^2}\widetilde{Q}(\lambda_{p_1})~\Sigma_{p_1p_1'}~\widetilde{Q}(\lambda_{p_1'}) =  \frac{u^2}{M^2}~\sum_{i,j}\sum_{p_2p_3p_4} \vp_i^{p_1}\vp_i^{p_2}\vp_i^{p_3}\vp_i^{p_4}\vp_j^{p_1'}\vp_j^{p_2}\vp_j^{p_3}\vp_j^{p_4}~ \widetilde{Q}(\lambda_{p_1})~\widetilde{Q}(\lambda_{p_1'})~\widetilde{Q}(\lambda_{p_2})~\widetilde{Q}(\lambda_{p_3})~\widetilde{Q}(\lambda_{p_4})
.
\eeq
Averaging over the eigenvalue distributions, the $p$ dependence of eigenvalues drop out and one can perform the summation over eigenvectors,
\beq
&&\frac{u^2}{M^2} ~\sum_{i,j}\sum_{p_2p_3p_4} \vp_i^{p_1}\vp_i^{p_2}\vp_i^{p_3}\vp_i^{p_4}\vp_j^{p_1'}\vp_j^{p_2}\vp_j^{p_3}\vp_j^{p_4}~\prod_{\{\lambda_{p_i}\}}\int d\lambda_{p_i}~ \widetilde{Q}(\lambda_{p_1})~ \widetilde{Q}(\lambda_{p_2})~\widetilde{Q}(\lambda_{p_3})~\widetilde{Q}(\lambda_{p_4})~\widetilde{Q}(\lambda_{p_5})~R_5(\lambda_{p_1},\lambda_{p_2},\lambda_{p_3},\lambda_{p_4},\lambda_{p_5})\\
&=&\frac{u^2}{M^2} \delta_{p_1,p_1'}~\prod_{\{\lambda_{p_i}\}}\int d\lambda_{p_i}~ \widetilde{Q}(\lambda_{p_1})~ \widetilde{Q}(\lambda_{p_2})~\widetilde{Q}(\lambda_{p_3})~\widetilde{Q}(\lambda_{p_4})~\widetilde{Q}(\lambda_{p_5})~R_5(\lambda_{p_1},\lambda_{p_2},\lambda_{p_3},\lambda_{p_4},\lambda_{p_5})
\eeq
where $R_5(...)$ is the 5-level correlation function in GOE. We can express the 5-level correlation function more generally as,
\beq
R_5(x_1,x_2,x_3,x_4,x_5) &=& \sigma_J(x_1)~\sigma_J(x_2)~\sigma_J(x_3)~\sigma_J(x_4)~\sigma_J(x_5)-\sum_P \sigma_J(x_{p_1})~\sigma_J(x_{p_2})~\sigma_J(x_{p_3})~T_2(x_{p_4},x_{p_5})\nonumber\\
&+&\sum_P \sigma_J(x_{p_1})~\sigma_J(x_{p_2})~T_3(x_{p_3},x_{p_4},x_{p_5})+\sum_P \sigma_J(x_{p_1})~T_2(x_{p_2},x_{p_3})T_2(x_{p_4},x_{p_5})\nonumber\\
&-&\sum_P \sigma_J(x_{p_1})~T_4(x_{p_2},x_{p_3},x_{p_4},x_{p_5})-\sum_P T_2(x_{p_1},x_{p_2})~T_3(x_{p_3},x_{p_4},x_{p_5})+T_5(x_{1},x_{2},x_{3},x_{4},x_{5}).
\label{5level}
\eeq 
The first term above is a simple (independent) product of the density of states for the five eigenvalues. The correlation among the different eigenvalues is contained in the $2,~3,~4$ and $5-$level ``cluster functions", $T_2,~T_3,~T_4,~T_5$, respectively \cite{RMTbook}. The summation $P$ is the permutation among $x_1$, $x_2$, $x_3$, $x_4$ and $x_5$. The first term in $R_5(x_1,..,x_5)$ above leads to the $\sim u^2|\omega|^5$ singular structure in the imaginary part of the self energy as in the main text since each integration $[\int~ d\lambda_i~ \sigma_J(\lambda_i) \widetilde{Q}(\lambda_i)]$ gives a factor of $Q(i\omega_n)$ (the rest follows the discussion in the main text). The remaining terms in Eqn.~\ref{5level}, that take into account correlations among eigenvalues of the random matrix, are not included in the replica treatment of the large$-N,~M$ saddle-point action for the rotor-theory.

Within the random-matrix picture, the higher-order corrections to the ``free" theory can be studied systematically by introducing the higher level correlation functions (i.e. $R_n(x_1,...,x_n)$) for GOE \cite{RMTbook}. As is clear from the above discussion, there is a piece for all $R_n(x_1,...,x_n)$ which corresponds simply to an independent product of the $\sigma_J(x_i)$; these are the terms that are also included in the replica action for the rotor theory.

\end{widetext}
\end{document}